\documentclass[preprint,showpacs,preprintnumbers,amsmath,amssymb]{revtex4}

\usepackage{amsmath}
\usepackage{latexsym}
\usepackage{amsfonts}

\begin{document}

\newcommand{\Zc}{\mbox{${\cal Z}_c$}}
\newcommand{\Nat}{{\cal I} \!\!\! {\cal N}}
\newcommand{\Ree}{{\Bbb R}}

\newcommand{\be}{\begin{eqnarray}}
\newcommand{\ee}{\end{eqnarray}}
\newcommand{\ba}{\begin{array}}
\newcommand{\ea}{\end{array}}
\newcommand{\bena}{\begin{eqnarray}}
\newcommand{\eena}{\end{eqnarray}}
\newcommand{\bdis}{\begin{displaymath}}	
\newcommand{\edis}{\end{displaymath}}
\newcommand{\bit}{\begin{itemize}}	
\newcommand{\eit}{\end{itemize}}
\newcommand{\ben}{\begin{enumerate}}	
\newcommand{\een}{\end{enumerate}}
\newcommand{\nid}{\noindent}
\newcommand{\cl}{\centerline}
\newcommand{\nl}{\newline}
\newcommand{\ul}{\underline}
\newcommand{\dd}{\quad}
\newcommand{\re}{{\cal I \! \!R}}
\newcommand{\co}{\,{ I \! \! \! \! C}}
\newcommand{\ze}{{\cal Z \! \! \! \!Z}}
\newcommand{\zp}{{\cal Z}(\beta)}
\newcommand{\zpn}{{\cal Z}(\beta, N)}
\newcommand{\zpc}{{\cal Z }_c ( \beta )}
\newcommand{\zpcn}{{\cal Z }_c ( \beta,N )}
\newcommand{\isn}{\Omega(v,N)}
\newcommand{\isnp}{\Omega^\prime(v,N)}
\newcommand{\isnpp}{\Omega^{\prime\prime}(v,N)}
\newcommand{\imp}{\int {\cal D }[q]}
\newcommand{\imf}{\int {\cal D }[\phi]}
\newcommand{\ebvq}{\exp \{ - \beta \sum_i  V[q_i] \}}
\newcommand{\ebvf}{\exp \{ - \beta \int dx   V[\phi(x)] \}}
\newcommand{\ebvfd}{\exp \{ - \beta \sum_i   V[\phi] \}}
\newcommand{\zee}{{\cal Z \! \! \! \!Z}_2}
\newcommand{\boh}{\Longleftrightarrow}
\newcommand{\void}{\lpg \o \rpg}
\newcommand{\mm}{{\rho_{\mu}}}
\newcommand{\mc}{{\rho_{can}}}
\newcommand{\mgc}{{\rho_{gc}}}
\newcommand{\ham}{{\cal H}}
\newcommand{\Xc}{{\cal X}}
\newcommand{\Yc}{{\cal Y}}
\newcommand{\Tc}{{\cal T}}
\newcommand{\Mc}{{\cal M}}
\newcommand{\Cc}{{\cal C}}
\newcommand{\vb}{\bar{v}}
\newcommand{\ub}{\bar{u}}
\newcommand{\vbx}{{v \! \! \! \!  -}_{max}}
\newcommand{\vbi}{{v \! \! \! \!  -}_{min}}
\newcommand{\Kc}{{\cal K}}
\newcommand{\Rc}{{\cal R}}
\newcommand{\Mcv}{{\cal M}_v}
\newcommand{\Mcg}{{\cal M}^G_g}
\newcommand{\Mce}{{\cal M}^{\cal H}_E}
\newcommand{\Si}{\Sigma}
\newcommand{\Siv}{\Sigma_v}
\newcommand{\Sig}{\Sigma^G_g}
\newcommand{\Sie}{\Sigma^{\cal H}_E}
\newcommand{\nG}{\frac{\partial^\mu G}{\| \nabla G \|}}
\newcommand{\nGn}{\frac{\partial^\mu G}{\| \nabla G \|^2}}
\newcommand{\nH}{\frac{\partial^\mu {\cal H}}{\| \nabla {\cal H} \|}}
\newcommand{\nHn}{\frac{\partial^\mu {\cal H}}{\| \nabla {\cal H} \|^2}}
\newcommand{\ngH}{\| \nabla {\cal H} \|}
\newcommand{\ngV}{\| \nabla V \|}
\newcommand{\gV}{ \nabla V}
\newcommand{\de}{\partial}
\newcommand{\too}{\longrightarrow}
\newcommand{\impl}{\Longrightarrow}
\newcommand{\eb}{{\bf e}}
\newcommand{\lpt}{\left(}
\newcommand{\rpt}{\right)}
\newcommand{\lpg}{\{}
\newcommand{\rpg}{\}}
\newcommand{\lpq}{\left[}
\newcommand{\rpq}{\right]}
\newcommand{\e}{\mbox{\boldmath $e$}}
\newcommand{\x}{\mbox{\boldmath $x$}}
\newcommand{\y}{\mbox{\boldmath $y$}}
\newcommand{\vi}{\mbox{\boldmath $i$}}
\newcommand{\vj}{\mbox{\boldmath $j$}}
\newcommand{\qs}{\frac{dq}{ds}}
\newcommand{\q}{\mbox{\boldmath $q$}}
\newcommand{\p}{\mbox{\boldmath $p$}}
\newcommand{\Q}{\mbox{\boldmath $Q$}}
\newcommand{\bP}{\mbox{\boldmath $P$}}
\newcommand{\bal}{\mbox{\boldmath $\alpha$}}
\newcommand{\A}{{\bf A}}
\newcommand{\bS}{{\bf S}}
\newcommand{\cn}{\mbox{cn}}
\newcommand{\sn}{\mbox{sn}}
\newcommand{\lps}{\langle}
\newcommand{\rps}{\rangle}
\newcommand{\dyy}{\displaystyle}
\newcommand{\nnv}{\frac{\nabla}{\| \nabla V \|}}
\newcommand{\unv}{\frac{1}{\| \nabla V \|}}
\newcommand{\ps}{\underline{\psi}}
\newcommand{\psv}{\underline{\psi}(V)}
\newcommand{\psc}{\underline{\psi}(\chi)}
\newcommand{\psvi}{{\psi_i}(V)}
\newcommand{\psvj}{{\psi_j}(V)}
\newcommand{\psvk}{{\psi_k}(V)}
\newcommand{\psvr}{{\psi_r}(V)}
\newcommand{\psci}{{\psi_i}(\chi)}
\newcommand{\pscj}{{\psi_j}(\chi)}
\newcommand{\psck}{{\psi_k}(\chi)}
\newcommand{\pscr}{{\psi_r}(\chi)}
\newcommand{\psvd}{\underline{\psi}(V) \cdot}
\newcommand{\pscd}{\underline{\psi}(\chi) \cdot}
\newcommand{\dpsv}{\underline{ \cdot \psi}(V)}
\newcommand{\dpsc}{\underline{ \cdot \psi}(\chi)}
\newcommand{\emme}{\nabla \frac{\nabla V}{\| \nabla V \|}}
\newcommand{\dvnv}{\frac{\triangle V}{\| \nabla V \|^2}}
\newcommand{\dv}{\triangle V}


\title{{A Theorem on the origin of Phase Transitions}}

\date{\today}

\author{Roberto Franzosi} 
\email{Roberto.Franzosi@df.unipi.it}
\affiliation{Dipartimento di Fisica Universit\`a di Pisa, and  I.N.F.N., 
Sezione di Pisa, and I.N.F.M., Unit\`a di Pisa, via Buonarroti 2, 
I-56127 Pisa, Italy }

\author{Marco Pettini}
\email{pettini@arcetri.astro.it}
\affiliation{Istituto Nazionale 
di Astrofisica, Largo E. Fermi 5,
 50125 Firenze, and I.N.F.M., Unit\`a di Firenze, and I.N.F.N., Sezione di 
Firenze, Italy  }
%



\begin{abstract}
For physical systems described by smooth, finite-range and confining 
microscopic interaction potentials $V$ with continuously varying coordinates, 
we announce and outline the proof of a theorem that establishes that unless 
the equipotential hypersurfaces of configuration space
$\Sigma_v =\{ (q_1,\dots,q_N)\in{\Bbb R}^N \vert\ V(q_1,\dots,q_N)=v\}$, 
$v\in{\Bbb R}$,
{\it change topology} at some $v_c$ in a given interval $[v_0, v_1]$ of 
values $v$ of $V$, the Helmoltz free energy must be at least twice 
differentiable
in the corresponding interval of inverse temperature $(\beta(v_0), \beta(v_1))$
also in the $N\rightarrow\infty$ limit. Thus the occurrence of a phase 
transition
at some $\beta_c =\beta(v_c)$ is {\it necessarily} the consequence of the loss 
of diffeomorphicity among the $\{\Sigma_v\}_{v < v_c}$ and the 
$\{\Sigma_v\}_{v > v_c}$, which is the consequence of the existence of
critical points of $V$ on $\Sigma_{v=v_c}$, that is points where $\nabla V=0$.
\end{abstract}
\pacs{ 05.70.Fh; 64.60.-i; 02.40.-k  }
\maketitle

Phase transitions (PTs) are phenomena which bring about {\it qualitative} 
physical
changes at the macroscopic level in presence of the same microscopic forces
acting among the constituents of a system. Their mathematical description
requires to translate into {\it quantitative} terms the mentioned qualitative 
changes. The standard way of doing this is to consider how the values of 
thermodynamic observables, obtained in laboratory experiments, vary with 
temperature, or volume, or an external field, and then to associate the 
experimentally observed discontinuities at a PT to the appearance of some kind 
of singularity entailing a loss of analyticity.
Despite the smoothness of the statistical measures, after the Yang-Lee theorem 
\cite{YLthm} we know that in the $N\rightarrow\infty$ limit non-analytic
behaviors of thermodynamic functions are possible whenever the analyticity
radius in the complex fugacity plane shrinks to zero, because this entails the 
loss of {\it uniform convergence} in $N$ (number of degrees of freedom) of any 
sequence of real-valued thermodynamic functions, and all this depends on the 
distribution of the zeros of the grand 
canonical partition function. Also the other developments of the rigorous 
theory of PTs, like that due to Dobrushin, Lanford and Ruelle on Gibbs 
measures \cite{georgii,ruelleTD}, identify PTs with the loss of
analyticity.

However, we can wonder whether this is the ultimate level of mathematical 
understanding of PT phenomena, or if some reduction to a more basic level is 
possible. 
The present paper addresses just this point and aims at providing a 
non-technical presentation of a new rigorous result, reported
in Ref. \cite{cmp}, making its conceptual meaning and prospective physical
interest accessible without going through the details of a lengthy mathematical
proof.
The new theorem says that non-analyticity
is the ``shadow'' of a more fundamental phenomenon occurring in configuration 
space: a {\it topology change} \cite{topch} within the family of equipotential 
hypersurfaces $\Sigma_v =\{ (q_1,\dots,q_N)\in{\Bbb R}^N \vert\ 
V(q_1, \dots ,q_N) = v\}$, where $V$ and $q_i$ are the microscopic interaction
potential and coordinates respectively. 
This topological approach to PTs stems from the numerical study of the 
Hamiltonian dynamical counterpart of phase transitions, and precisely from the
observation of discontinuous or cuspy patterns displayed by the largest 
Lyapunov exponent at the transition energy \cite{physrep} (or temperature). 
Lyapunov exponents measure the strength of dynamical chaos and cannot be 
measured in laboratory experiments, at variance with thermodynamic observables,
thus, being genuine dynamical observables they are only measurable in 
numerical simulations of the microscopic dynamics. To get a hold
of the reason why the  largest Lyapunov exponent $\lambda_1$ should probe 
configuration space topology,
let us first remember that for standard Hamiltonian systems, described by
$H=\sum_{i=1}^N\frac{1}{2}p_i^2 + V(q_1,\dots,q_N)$, $\lambda_1$ is 
computed by  solving the tangent dynamics equation
\begin{equation}
\frac{d^2\xi_i}{dt^2}+ \left(\frac{\partial^2 V}{\partial q^i\partial q^j}
\right)_{q(t)} \xi^j = 0~,
\label{tg-dyn}
\end{equation}
where $q(t)=[q_1(t),..,q_N(t)]$, and then $\lambda_1=\lim_{t\to\infty}1/2t \log
 (\Sigma_{i=1}^N[\dot\xi_i^2(t)+\xi_i^2(t)] / \Sigma_{i=1}^N[\dot\xi_i^2(0)+
\xi_i^2(0)] )$.
If there are critical points of $V$ in configuration space, that is points
$q_c=[{\overline q}_1,\dots,{\overline q}_N]$ such that 
$\left.\nabla V(q)\right\vert_{q=q_c}=0$, according to the 
{\it Morse Lemma} \cite{hirsch}, 
in the neighborhood of any critical point $q_c$ there always exists a 
coordinate system ${\tilde q}(t)=[{\tilde q}_1(t),..,{\tilde q}_N(t)]$ 
for which
\begin{equation}
V({\tilde q})= V(q_c) - {\tilde q}_1^2-\dots -{\tilde q}_k^2 + 
{\tilde q}_{k+1}^2+\dots +{\tilde q}_N^2~,
\label{morsechart}
\end{equation}
where $k$ is the index of the critical point, i.e. the number of negative
eigenvalues of the Hessian of $V$.
In the neighborhood of a critical point, Eq.(\ref{morsechart}) yields  
$\partial^2_{ij} V =\pm \delta_{ij}$ which, substituted into 
Eq.(\ref{tg-dyn}), gives $k$ unstable directions which contribute \cite{nota} 
to the exponential growth of the norm of the tangent vector $\xi$. 
This means that 
the strength of dynamical chaos, measured by the largest Lyapunov 
exponent $\lambda_1$, is affected by the existence of critical points of $V$. 
In particular, let us consider the possibility of a sudden variation, with
the potential energy $v$, of the number of critical points (or of their 
indexes) in configuration space at some value $v_c$, it is then reasonable to
expect that the pattern of $\lambda_1(v)$ -- as well as that of $\lambda_1(E)$
since $v=v(E)$ --  will be consequently affected, thus displaying jumps or
cusps or other ``singular'' patterns at $v_c$ (this heuristic argument has
been given evidence in the case of the XY-mean-field model, see \cite{physrep} 
and \cite{xymf}).  
On the other hand, {\it Morse theory} \cite{hirsch} teaches us that the 
existence of critical points of $V$ is associated with topology changes of the
hypersurfaces $\{\Sigma_v\}_{v\in{\Bbb R}}$, provided that $V$ is a good Morse
function (that is: bounded below, with no vanishing eigenvalues of its Hessian
matrix). Thus the existence of critical points of the potential $V$ makes 
possible a conceptual link between dynamics and configuration space topology,
which, on the basis of both direct and indirect
evidence for a few particular models, has been formulated \cite{physrep} as 
a {\it topological hypothesis} about the relevance of topology for PTs
phenomena.
 In what follows, we show that, for a large class of physically
meaningful potentials, this conjectural status of the art turns into a
qualitatively new one because we can prove the following
\vskip 0.2truecm
\noindent{\bf Theorem.} {\it Let $V_N(q_1,\dots,q_N): {\Bbb R}^N \rightarrow 
{\Bbb R}$, be a smooth, bounded from below, finite-range and confining 
potential \cite{notap}. Denote by $\Sigma_v:= V^{-1}(v)$, $v\in{\Bbb R}$, its  
{\em level sets}, or {\em equipotential hypersurfaces}, in configuration space.
Then let $\vb =v/N$ be the potential energy per degree of freedom.

If there exists $N_0$, and if for any pair of values $\vb$ and $\vb^\prime$ 
belonging  to a given
interval $I_{\vb}=[\vb_0, \vb_1]$ and for any $N>N_0$ 

\centerline{$\Sigma_{N\vb}$ is {\em diffeomorphic} to $\Sigma_{N\vb^\prime}$}
\vskip 0.2truecm
then the sequence of the Helmoltz free energies 
$\{ F_N (\beta)\}_{N\in{\Bbb N}}$ -- where $\beta =1/T$ ($T$ is the 
temperature) and 
$\beta\in I_\beta =(\beta (\vb_0), \beta (\vb_1))$ -- is {\em uniformly}
convergent at least in ${\cal C}^2(I_\beta)$ [the space of twice 
differentiable functions in the interval $I_\beta$], so that 
$\lim_{N\to\infty}F_N \in{\cal C}^2(I_\beta)$ and neither first nor second 
order phase transitions can occur in the (inverse) temperature interval 
$(\beta (\vb_0), \beta (\vb_1))$. }
\smallskip

\noindent Where the inverse temperature is defined as 
$\beta (\vb) =\partial S_N^{(-)}(\vb)/\partial \vb$ and
$S_N^{(-)}(\vb)=N^{-1}\log\int_{V(q)\le \vb N}\ d^Nq$ is one of the possible
definitions of the microcanonical configurational entropy. The intensive
variable $\vb$ has been introduced to ease the comparison between quantities
computed at different $N$-values.
\smallskip\smallskip

\noindent This theorem means that a topology change of the 
$\{ \Sigma_v\}_{v\in{\Bbb R}}$ at some $v_c$ is a {\it necessary} 
condition for a phase transition to take place at the corresponding energy or 
temperature value. The topology changes implied here are those described 
within the framework of Morse theory through {\it attachment of handles} 
\cite{hirsch,notatop}.
\smallskip

\noindent{\bf Remark 1. }{\it The topological condition of diffeomorphicity 
among all the hypersurfaces $\Sigma_{N\vb}$ with $\vb\in[\vb_0, \vb_1]$
has an  analytical consequence: the absence of 
critical points of $V$ in the interval $[\vb_0, \vb_1]$. This is proved
in Lemma 1 of Ref.{\rm\cite{cmp}} by 
adapting to the $\Sigma_v$ Bott's ``critical neck theorem''{\rm\cite{notatop}}
which applies to the manifolds $M_v=\{(q_1,...,q_N)\in{\Bbb R}^N\vert
V(q_1,...,q_N)\leq v\}$.
Apart from 
this initial link with topology, the proof proceeds in the domain of Analysis.}
\smallskip

\noindent{\bf Remark 2. }{\it In the proof we resort to the concept of
{\rm uniform
convergence} -- from elementary functional analysis -- of a sequence of 
functions,
and to the fact that the limit of a sequence of smooth functions can be
non-smooth.
This way of tackling the thermodynamic limit is in the spirit of the 
celebrated Yang-Lee theorem \rm{\cite{YLthm}}.}
\smallskip

Let us now outline the proof by focusing on the main ideas (details can be 
found in \cite{cmp}).

Under the {\it crucial hypothesis of diffeomorphicity} of the hypersurfaces 
$\Sigma_{N\vb}$ for $\vb\in [\vb_0, \vb_1]$, we want to prove that the 
thermodynamic limit of the Helmoltz free energy, $F_\infty (\beta)=
\lim_{N\to\infty} F_N(\beta)$, is at
least twice differentiable,  so that first or second order phase transitions 
are absent.
For standard Hamiltonians, each function 
$F_N(\beta)$ reads as $F_N(\beta)=-(2\beta)^{-1}\log(\pi/\beta)- f_N(\beta)/
\beta$, sum of a part coming from the kinetic energy term, and a 
configurational part $f_N(\beta )=(1/N)\log \int d^N q\ \exp [-\beta V(q)]$. 
Thus, in order to prove that $F_\infty (\beta)\in{\cal C}^2(I_\beta)$, 
we have to show that the sequence of smooth functions
$\{F_N(\beta)\}_{N\in{\Bbb N}}$ is {\it uniformly convergent} at least in 
${\cal C}^2(I_\beta)$ in the limit $N\rightarrow\infty$, or equivalently,
since $(2\beta)^{-1}\log(\pi/\beta)$ remains always smooth in  the limit 
$N\to\infty$, we have to show that the sequence of smooth functions 
$\{f_N(\beta)\}_{N\in{\Bbb N}}$ is uniformly convergent in 
${\cal C}^2(I_\beta)$ when $N\rightarrow\infty$.
Now, at any $N$, $f_N(\beta )$ is related to the microcanonical entropy
$S_N^{(-)}$ through the Legendre transform:
$S^{(-)}_N(\vb) = f_N(\beta) + \beta \vb$. 
Hence, to prove that $f_\infty (\beta)$ is twice 
differentiable with respect to $\beta$ we need to prove that 
$S^{(-)}_\infty (\vb)$ is three times differentiable with respect to $\vb$. 

Eventually, we consider the
equivalent definition, at large $N$, of the configurational microcanonical 
entropy 
\begin{equation}
S_N(v)= \frac{1}{N} \log \ \Omega(v, N)\equiv
 \frac{1}{N} \log\int_{\Sigma_v} \frac{d\sigma}{\Vert \nabla V\Vert}~,
\label{Sconf}
\end{equation}
which also implicitly defines $\Omega$ as the surface integral in the r.h.s.,
where $d\sigma$ is the $(N-1)$-dimensional surface element of $\Sigma_v$, and
where $\Vert \nabla V\Vert =[\sum_{i=1}^N(\partial V/\partial q_i)^2]^{1/2}$; 
since $S_N(\vb)$ has the same thermodynamic limit of the entropy 
$S^{(-)}_N(\vb)$, that is $S^{(-)}_\infty (\vb )=S_\infty (\vb )$,  
we are left with the problem of
proving that the sequence of smooth functions
$\{S_N(\vb)\}_{N\in{\Bbb N}}$ [where $S_N(v)=S_N(\vb N)$], 
is uniformly convergent in ${\cal C}^3(I_{\vb})$,
the space of three times differentiable functions in the interval $I_{\vb}$,  
in the limit $N\rightarrow\infty$. The reason for using $S_N(\vb)$ instead of
$S^{(-)}_N(\vb)$ will be soon clear.
After the Ascoli theorem \cite{schwartz}, in order 
to prove that $S_\infty (\vb)$ is three times differentiable, we need to prove 
that for $\vb\in I_{\vb}=[\vb_0, \vb_1]$ 
and {\it for any} $N$, the function
$S_N(\vb)$ and its first four derivatives are uniformly bounded in $N$ from 
above, that is, for any $N\in{\Bbb N}$ and $\vb \in [\vb_0, \vb_1 ]$
\begin{equation}
	\sup \left\vert S_N({\vb})\right\vert
        < \infty \, , \, 
 ~\sup\left\vert \frac{\de^k
	S_N}{\de {\vb}^k}
\right\vert < \infty\, ,~k=1,..,4 . 
\label{bounds}
\end{equation}

We prove the Theorem by proving that these bounds are the consequence of the 
diffeomorphicity among the $\Sigma_{N\vb}$, for $\vb\in [\vb_0, \vb_1]$. 

From Eq.(\ref{Sconf}) the first four derivatives of $S_N(\vb)$ are trivially 
computed:
	\bena
	\frac{\de 
	S_N}{\de {\vb}}({\vb}) 
	= \frac{1}{N} \frac{\isnp}{\isn} \cdot \frac{d v}{d {\vb}} =
	\frac{\isnp}{\isn} \label{ds1}
	\eena
and, using a compact notation, 
$\partial^2_{\vb}S_N=N[\Omega^{\prime\prime}/\Omega -
(\Omega^\prime/\Omega)^2]$, $\partial^3_{\vb}S_N=N^2
[\Omega^{\prime\prime\prime}/\Omega -
3\Omega^{\prime\prime}\Omega^\prime/\Omega^2
+2(\Omega^\prime/\Omega)^3]$ and
$\partial^4_{\vb}S_N=N^3[ \Omega^{iv}/\Omega -4\Omega^{\prime\prime\prime}
\Omega^\prime/\Omega^2 -3(\Omega^{\prime\prime}/\Omega )^2 +12
\Omega^{\prime\prime}(\Omega^\prime)^2/\Omega^3
-6(\Omega^\prime/\Omega)^4]$, where the prime indexes stand 
for derivations of $\Omega(v,N)$ with respect to $v=\vb N$.
In order to verify whether the conditions  (\ref{bounds}) are fulfilled, we
must be able to estimate the $N$-dependence of all the addenda in these
expressions for the derivatives of $S_N$.

Being the assumption of diffeomorphicity of the  $\Sigma_{N\vb}$ equivalent 
to the absence of critical points of the potential, we can 
use the derivation formula \cite{federer,laurence}
	\bena	
\frac{d^k}{dv^k}\isn = \int_{\Si_v}\ngV \ A^k \lpt \unv \rpt \frac{d \sigma}
                       {\ngV} \dd ,
	\label{isnp}
	\eena
where $A^k$ stands for $k$ iterations of the operator
\[
A(\bullet ) =\nabla \lpt \frac{\nabla V}{\| \nabla V \|}\ \bullet \rpt \frac{1}
{\| \nabla V \|}~.
\]
The technical reason to work with $S_N$ instead of $S_N^{(-)}$ is now evident:
the derivatives of $\Omega (v,N)$ are transformed into the surface integrals
of explicitly computable combinations and powers of a few basic ingredients, 
like 
$\Vert\nabla V\Vert$, $\partial V/\partial q_i$, 
$\partial^2 V/\partial q_i\partial q_j$, 
$\partial^3 V/\partial q_i\partial q_j\partial q_k$ and so on.
This is a technically crucial step to prove the Theorem.

The first uniform bound in Eq.(\ref{bounds}), $\vert S_N({\vb})\vert < \infty$,
is a simple consequence of the intensivity of $S_N({\vb})$.

To prove the boundedness of the first derivative of $S_N$, we first compute 
its expression by means of Eqs.(\ref{ds1}) and (\ref{isnp}), which reads
\begin{equation}
\frac{\de S_N}{\de {\vb}}=\frac{1}{\Omega}\int_{\Sigma_{\vb N}}\left[
\frac{\Delta V}{\ngV^2}
  -2 \frac{\sum_{i,j}
\partial^i V\partial^2_{ij} V \partial^j V}{\ngV^4}\right]
\frac{d\sigma}{\ngV}~,
\label{der1}
\end{equation}
with $\partial_iV=\partial V/\partial q^i$ and $i,j=1,\dots,N$, whence 
(with an obvious meaning of $\langle\cdot\rangle_{\Sigma_v}$)  
\begin{equation}
\left\vert\frac{\de S_N}{\de {\vb}}\right\vert \leq
\left \langle \frac{\mid \Delta V \mid}{ \ngV^2}\right \rangle_{\Sigma_v}
  +2\left \langle \frac{\left\vert \sum_{i,j}
\partial^i V\partial^2_{ij} V \partial^j V 
\right \vert }{\ngV^4}\right \rangle_{\Sigma_v}.
\label{ineq}
\end{equation}
The $N$-dependences of the derivatives of $S_N$ are estimated at constant
potential energy density $\vb$, for any given $\vb\in [\vb_0,\vb_1]$, thus 
we can think of increasing $N$ by glueing together an increasing number $k$ 
of replicas of 
a given building block of $N_0$ particles at potential energy $v=\vb N_0$.
Now the {\it hypothesis of diffeomorphicity} of the $\Sigma_v$ plays again a
crucial role, in fact {\it in the absence of critical points}, for each 
building block we have
$\min \Vert\nabla V_{N_0}\Vert^2 \geq {\bar C}^2 >0$, with ${\bar C}$ a 
constant, and as 
only short-range potentials are considered, the larger $N_0$ and $N (=kN_0)$ 
the  less relevant are the boundary interactions among the building blocks. 
Thus, at large $N$,
$\min \ngV^2 \geq C^2 N$, where $C={\bar C}/N_0$ is a constant; 
for an upper bound estimate of Eq.(\ref{ineq})
we replace in its denominators the lower bound $C^2 N$ of $\min \ngV^2$
\[
\left\vert\frac{\de S_N}{\de {\vb}}\right\vert \leq
\frac{\left \langle \mid \Delta V \mid \right \rangle_{\Sigma_v}}{C^2~N} 
+2 \frac{\left \langle  \left\vert \sum_{i,j}
\partial^i V\partial^2_{ij} V \partial^j V \right\vert 
\right \rangle_{\Sigma_v}}{C^4~N^2}~,
\]
where now we have to estimate the $N$-dependence of the numerators. To this
purpose, as we have assumed that $V$ is smooth and bounded below, we note 
that $\langle \mid \Delta V \mid \rangle_{\Sigma_v}
=\langle \mid \sum_{i=1}^N\partial^2_{ii} V \mid \rangle_{\Sigma_v}
\leq N\max_i \langle  \mid \partial_{ii}^2 V \mid 
 \rangle_{\Sigma_v}$ and, as we have also assumed that $V$ is a short range 
potential, the number of non-vanishing matrix elements $\partial^2_{ij}V$ is 
$N(d+1)$ where  $d$ is the number of neighbouring particles in the
interaction range of the potential, thus 
$\left \langle \mid \partial^i V\partial^2_{ij} V \partial^j V
\mid\right\rangle_{\Sigma_v} \leq  N (d+1)\max_{i,j} \langle \mid \partial^i V
\partial^2_{ij} V \partial^j V \mid\rangle_{\Sigma_v}$. 
Finally, putting $ m=\max_{i,j} \langle \mid \partial^i
 V \partial^2_{ij} V \partial^j V \mid\rangle_{\Sigma_v}$
\begin{equation}
\left\vert\frac{\de S_N}{\de {\vb}}\right\vert\leq 
\frac{\max_i \langle  \mid \partial_{ii}^2 V \mid 
 \rangle_{\Sigma_v}}{C^2} + 2\frac{ m (d+1)}{C^4 N}
\label{maggior1}
\end{equation}
which, in the limit $N\rightarrow\infty$, shows 
that the first derivative of the entropy is uniformly
bounded by a finite constant. This first step proves that $S_\infty (\vb )$ 
is continuous.

The three further steps, concerning boundedness of the higher order 
derivatives, involve similar arguments to be applied to a number of terms 
which is rapidly increasing with the order of the derivative.
But many of these terms can be grouped in the form of the variance or
higher moments of certain quantities, thus allowing the use of a powerful 
technical trick to compute their $N$-dependence. 
For example, using Eq.(\ref{isnp}) in
the expression for $\partial^2_{\vb}S_N$ just below Eq.(\ref{ds1}), we get 
 \be
  \left | \frac{\de^2 S_{N}}{\de {\vb}^2} \right |
  \leq N \Big |
  \langle\alpha ^2\rangle_{\Sigma_v}\!
  - \langle\alpha\rangle_{\Sigma_v}^2\Big | 
  + N \Big |  \langle
  \psvd \ps \lpt \alpha \rpt
  \rangle_{\Sigma_v} \Big |
  \label{f2v_3}
  \ee
where $\alpha ={\ngV\ A (1/\ngV )}$ and $\ps =\nabla/\ngV$. Now, 
it is possible to think of the scalar function $\alpha$ as if it were
a random variable, so that
the first term in the r.h.s. of Eq.(\ref{f2v_3}) would be its second moment.
Such a possibility is related with the general validity of the Monte Carlo
method to compute multiple integrals.
In particular, since the $\Sigma_v$ are smooth, closed ($V$ is non-singular),
without critical points and representable as the union of suitable subsets
of ${\Bbb R}^{N-1}$, the standard Monte Carlo method  \cite{mcmc}
is applicable to 
the computation of the averages $\langle\cdot\rangle_{\Sigma_v}$ which become
sums of standard integrals in ${\Bbb R}^{N-1}$. This means
that a random walk can be 
constructively defined on any $\Sigma_v$, which conveniently samples the 
desired measure on the surface. 
Along such a random walk, usually called Monte Carlo Markov 
Chain (MCMC), $\alpha$ and its powers
behave as random variables whose ``time'' averages along the MCMC converge to
the surface averages $\langle\cdot\rangle_{\Sigma_v}$.
Notice that the actual computation of these surface averages goes beyond 
our aim, in fact, we do not need the numerical values -- but only the 
$N$-dependences -- of the upper bounds of the derivatives of the entropy. 
Therefore, all what we need is just knowing that in principle a suitable MCMC
exists on each $\Sigma_v$.
Now, the function $\alpha$ is the integrand in square brackets in 
Eq.(\ref{der1}), where the second term vanishes at large $N$, as is clear from
Eq.(\ref{maggior1}). Therefore, at increasingly large $N$, the approximate
 expression 
$\alpha =\sum_{i=1}^N \partial^2_{ii}V/\Vert\nabla V\Vert^2$ tends to become 
exact. $\alpha$ is in the form of a sum function
$\alpha  =N^{-1} \sum_{i =1}^N a_i$ of terms 
$a_i=N \partial^2_{ii}V/\Vert\nabla V\Vert^2$, of ${\cal O}(1)$ in $N$, which, 
along a MCMC, behave as
independent random variables with probability densities $u_i (a_i)$ which 
we do not need to know explicitly. 
Then, after a classical ergodic theorem for sum functions, due to Khinchin 
\cite{khinchin}, 
based on the Central Limit Theorem of probability theory, $\alpha$ is a
gaussian-distributed random variable; as its variance decreases linearly 
with $N$, 
 $\lim_{N\to\infty}  N  | \langle {\alpha}^2\rangle_{\Sigma_v}\!
-  \langle {\alpha} \rangle_{\Sigma_v}^2 | = const <\infty$.

Arguments similar to those above used for the first derivative of $S_N$
lead to the result
$\lim_{N\to\infty}  N  |  \langle\psvd \ps \lpt \alpha \rpt\rangle_{\Sigma_v}  
|= const <\infty$, which, together with what has been just found for 
the variance 
of $\alpha$, proves the uniform boundedness also of the second derivative of 
$S_N$ under the hypothesis of diffeomorphicity of the $\Sigma_v$.

Similarly, but with an increasingly tedious work, we can treat the 
third and fourth derivatives of the entropy. In fact, despite the large number
of terms contained in their expressions, they again belong only to two 
different categories: those terms which can be grouped in the form of higher 
moments of the function $\alpha$, and whose $N$-dependence is known after
the above mentioned theorem due to Khinchin, and those terms whose 
$N$-dependence can be found by means of the same kind of estimates 
given above for $\partial_{\vb}S_N$. Eventually, after a lenghty but rather
mechanical work, also the third and fourth derivatives of $S_N$ are shown
to be uniformly bounded as prescribed by Eq.(\ref{bounds}). 
Whence the proof of the Theorem.
 
A few comments are in order. 

The converse of our Theorem is not true. 
There is not a one-to-one correspondence 
between phase transitions and topology changes, 
in fact, there are smooth, confining and finite-range potentials, like the 
one-dimensional XY model \cite{xymf}, 
with even a very large number of critical 
points, and thus many changes in the topology of the $\Sigma_v$, 
but with no phase transition.
Therefore, 
an open problem is that of {\it sufficiency} conditions, that is to determine
which kinds of topology changes can entail the appearance of a PT. 
Preliminary hints on this point are given by 
the analytic study of particular models \cite{xymf,ptrig} for which topology 
and thermodynamics are exactly computed.

Finally, though at present our Theorem only applies to first and second order 
PTs and to those systems for which $V$ is a good Morse function, it provides 
the grounding to an approach which can unify the mathematical 
description of very different kinds of PTs, like those ``exotic'' ones 
occurring in glasses or in the folding of polymers and proteins, for which
the so-called {\it energy landscape paradigm} \cite{elandscape} 
is currently
studied overlooking the link with Morse theory and topology.

We thank  H. van Beijeren, L. Casetti, E. Guadagnini, R. Lima, C. Liverani, 
A. Moro, P. Picco, M. Rasetti and G. Vezzosi for discussions and comments.



\begin{thebibliography}{99}
\bibitem{topch} Two given manifolds $\Sigma_v$ and $\Sigma_{v^\prime}$ have 
the same topology if they can be continuously and differentiably deformed one 
into the other, that is if they are diffeomorphic. 
Thus by ``topology change'' we mean ``loss of diffeomorphicity''.
\bibitem{nota} ``contribute'' because chaos stems also from a different
mechanism as is discussed in Ref.\protect\cite{physrep}.
\bibitem{notap} These requirements for $V$ are fulfilled by standard 
interatomic and intermolecular interaction potentials, as well as by classical 
spin potentials.
\end{thebibliography}
\end{document}